\documentclass[twocolumn]{jpsj2}
\newcommand{\be}{\begin{equation}}
\newcommand{\ee}{\end{equation}}
\newcommand{\bea}{\begin{eqnarray}}
\newcommand{\eea}{\end{eqnarray}}

\newcommand{\ket}[1]{\left\vert {#1} \right\rangle}

\newcommand{\aver}[1]{\left\langle {#1} \right\rangle}
\def\dfrac#1#2{{\displaystyle\frac{#1}{#2}}}
\def\H{{\cal H}}
\def\v#1{\mib #1}
\def\c{{\rm{c}}}
\def\const{{{\rm const}}}
\def\uud{\downarrow\uparrow\uparrow\downarrow\uparrow\uparrow...\ }
\def\udu{\uparrow\uparrow\downarrow\uparrow\uparrow\downarrow...\ }
\def\duu{\uparrow\downarrow\uparrow\uparrow\downarrow\uparrow...\ }
\def\sing{\mbox{$\bullet$\hskip -1pt-\hskip -1pt$\bullet $}}
\def\0u0{\mbox{$\bullet$\hskip -1pt-\hskip -1pt$\bullet\uparrow\bullet$\mbox{\hskip -1pt-\hskip -1pt}$\bullet\uparrow...$\ }}
\def\uudtype{{\uparrow\downarrow\uparrow\ }}
\def\u00type{\mbox{$\bullet$\hskip -1pt-\hskip -1pt$\bullet\uparrow$}}

\def\runauthor{Kazuo {\sc Hida} and Ian {\sc Affleck}}

\title{Quantum vs. Classical Magnetization Plateaus of S=1/2  Frustrated Heisenberg Chains }

\author
{Kazuo {\textsc Hida}$^{1}$
\thanks{E-mail: hida@phy.saitama-u.ac.jp} and Ian {\textsc Affleck}$^{2}$\thanks{E-mail: iaffleck@physics.ubc.ca}}

\inst
{$^{1}$Department of Physics, Faculty of Science,\\ Saitama University, Saitama, Saitama 338-8570\\
$^{2}$Department of Physics and Astronomy,\\ University of British Columbia, Vancouver, B.C., Canada V6T 1Z1}

\recdate
{\today
}

\abst
{
The competition between quantum and classical magnetization plateaus of $S=1/2$ frustrated Heisenberg chains with modified exchange couplings is investigated. The conventional $S=1/2$ frustrated Heisenberg chain is known to exhibit a 3-fold degenerate $\uudtype$-type classical plateau at 1/3 of the saturation magnetization accompanied by the spontaneous $Z_3$ translational symmetry breakdown. The stability of this plateau phase against period 3 exchange modulation which favors the $\u00type$-type quantum plateau state ($\sing = $  singlet dimer) is studied  by bosonization, renormalization group and numerical diagonalization methods. The ground state phase diagram and the spin configuration in each phase are  numerically determined. The {\it translationally invariant} 
Valence Bond Solid-type model with 4-spin and third neighbor interactions, which has the exact $\u00type$-type quantum plateau state, is also presented. The phase transition to the classical $\uudtype$-type ground state is also observed by varying the strength of 4-spin and third neighbor interactions. The relation between these two types of models with quantum plateau states is discussed.
} 
\kword
{
frustrated Heisenberg chain, magnetization plateau, period 3 modulation, 4-spin interaction, $Z_2$ symmetry breakdown, bosonization, numerical diagonalization
}
\begin{document}
\sloppy
\maketitle

\section{Introduction}
The phenomenon of magnetization plateau has been extensively studied as a macroscopic manifestation of the essentially quantum effect in which the magnetization $M$ is quantized at fractional values of the saturation magnetization $M_s$ in low dimensional magnetism\cite{kh,cg,oya,ok,ko,ok2,oku,oku2,tone}. 

\begin{figure}
\centerline{\includegraphics[width=60mm]{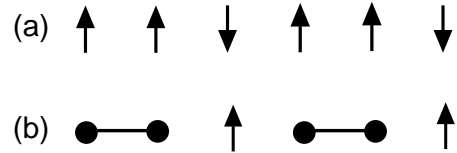}}
\caption{Two possible spin configurations in the $m=1/3$ plateau state. The classical configuration, (a), 
 is actually observed in the model (\ref{model1}) and the quantum configuration, (b), 
 in the model (\ref{model2}) for $\alpha >0$, with period 3 exchange modulation
 and also in the translationally invariant model (\ref{fourspin}). }
\label{fig1}
\end{figure}

There are at least two types of plateau states that can occur for magnetization $m=M/M_s=1/3$ in $S=1/2$ chain models. 
One is the ``classical state'' of Fig. \ref{fig1}(a), which occurs as an adiabatic continuation from the 
Ising limit.  The other is the $\0u0$ spin configuration shown in  Fig. \ref{fig1}(b) where $\sing$ stands for the singlet pair. This plateau is essentially of quantum origin, driven by the tendency of neighboring antiferromagnetically coupled spins to form entangled singlet states.\
  Since these are, in general, only simple caricatures of the ground states, we might ask whether they really correspond to distinct phases.  
 The ground state symmetries are essentially the same in both phases. The unit cell consists of 3 sites, as required 
by the Oshikawa-Yamanaka-Affleck theorem,\cite{oya} and there are reflection symmetries about every third site and every third link in both cases. 
To see that, nonetheless,  that they really 
{\it are} distinct phases note that in the classical state the local spin expectation value  $\aver{S^z_i}$ is larger on 2 out of 3 sites in a unit cell.  On the other hand, in the quantum state, it is smaller on 2 out of 3 sites, the sites where the dimers are drawn.  Clearly this feature implies distinct phases. 

Recently, Okunishi and Tonegawa\cite{oku,oku2} and Tonegawa and coworkers\cite{tone} investigated the magnetization process of the strongly frustrated Heisenberg chain, 
\begin{eqnarray}
\label{model1}
{\cal H}&=&J\sum_{i=1}^{L} \left[\v{S}_{i}\v{S}_{i+1} +\delta\v{S}_{i}\v{S}_{i+2}\right],
\end{eqnarray}
where $\v{S}_{i}$ is the spin-1/2 operator on the $i$-th site. In what follows, we take $J=1$ to fix the energy unit and assume the periodic boundary condition $\v{S}_{L+1}=\v{S}_{1}$ unless specifically mentioned. These authors pointed out that this model has a classical plateau at $m=M/M_s=1/3$ for $\delta >\delta_{\c} \equiv 0.487$.\cite{tone}

On the other hand, if we allow for the 4-spin interaction, we can also construct a 
 model which has exactly the quantum plateau by the Valence-Bond-Solid\cite{oya,aklt,mg}-type construction as follows,
\begin{eqnarray}
\H&=& \frac{1}{24}\sum_{i=1}^L \v{T}_i^2(\v{T}_i^2-2)
\label{fourspin}
\end{eqnarray}
where $\v{T}_i=\v{S}_i+\v{S}_{i+1}+\v{S}_{i+2}+\v{S}_{i+3}$. Each term of this Hamiltonian is non-negative definite and projects out the 4-spin singlet ($T=0$) and triplet states ($T=1$) where $T(T+1)=\v{T_i}^2$. Therefore if we arrange the dimers to sit on every third bond as shown in Fig. \ref{fig1}(b), we obtain a ground state of this Hamiltonian, since the sum of any 4 neighboring spins has no projection onto $T=2$ states in this configuration.  

In the plateau state, one third of the spins remain undimerized and they can be in an arbitrary state in zero field.  These spins are essentially paramagnetic and get polarized by an infinitesimal field to yield the 1/3 plateau state. Although this ground state, at zero field, is macroscopically  $2^{L/3}$-fold degenerate, this degeneracy is lifted by any small perturbation. For example, an additional small antiferromagnetic third neighbor antiferromagnetic interaction leads to the Tomonaga-Luttinger liquid ground state in the absence of a magnetic field. These weakly coupled spin are easily magnetized up to the $m=1/3$ plateau state. Conversely, a small ferromagnetic third neighbor interaction leads to ferrimagnetism (i.e. a moment of $m=1/3$ at zero field). 

It is also interesting to interpret these two types of plateau states in terms of the  Jordan-Wigner fermions. In the fermionic language, the classical plateau state corresponds to the usual site-centered CDW phase in which particles sit on two successive sites and every third site is empty, while the quantum plateau phase is a sort of a bond-centered CDW or 'bond order wave' (BOW) state where the fermion is not localized on a single site but tunnels back and forth between a pair of sites and the third site is filled. In this context, the distinction between classical and quantum plateau states is reminiscent of the distinction between CDW and BOW states in itenerant electronic system both of which seem to occur in the phase diagram of the half-filled extended Hubbard model in one dimension and related models.\cite{sbc} Considering the difference of the nature of these two types of plateau states, it is an interesting issue how these two plateau states transform into each other.

The quantum plateau state was actually 
first discovered, over a decade ago in the ferromagnetic-ferromagnetic-antiferromagetic 
(F-F-AF) chain model.\cite{kh,ko,ok,ok2} This is a special case of a more general class 
of models with period 3 space inversion  invariant exchange modulation:
\begin{eqnarray}
\label{model2}
{\cal H}&=&\sum_{l=1}^{L/3} \left[(1-\alpha)\left(\v{S}_{3l-1}\v{S}_{3l} +\v{S}_{3l}\v{S}_{3l+1}\right)\right.\nonumber \\
&+&\left.(1+\alpha)\v{S}_{3l+1}\v{S}_{3l+2}\right].
\end{eqnarray}
For $\alpha >1$ we obtain the F-F-AF model. The dimers now form uniquely on the antiferromagetic links.  
 In this case, of course, there is no spontaneous breaking of translational symmetry since it is already explicitly broken by the Hamiltonian.  A more general model with arbitrary periodicity is investigated by Cabra and Grynberg.\cite{cg}

The competition between quantum and classical plateaus is naively realized in the  frustrated Heisenberg chain with period 3 space inversion  invariant exchange modulation described by the Hamiltonian,
\begin{eqnarray}
\label{model3}
{\cal H} &=&\sum_{l=1}^{L/3} \left[(1-\alpha)\left(\v{S}_{3l-1}\v{S}_{3l} +\v{S}_{3l}\v{S}_{3l+1}\right)\right.\nonumber \\
&+&\left.(1+\alpha)\v{S}_{3l+1}\v{S}_{3l+2}\right]+\delta\sum_{i=1}^{L}\v{S}_{i}\v{S}_{i+2}.
\end{eqnarray}
It is obvious that this model shows a quantum plateau for large enough positive $\alpha$ and a classical plateau for $\alpha=0$ and large enough $\delta$.

This paper is organized as follows. In the next section, the low energy effective boson theory for the $m=1/3$ plateau state of the translationally invariant frustrated Heisenberg model is presented. The effect of the period 3 exchange modulation is analyzed using the renormalization group method in \S 3. The quantitative ground state phase diagram at $m=1/3$ is obtained by the numerical diagonalization method in \S 4. The nature of each phase is also discussed. In \S 5, the VBS-type model with 4-spin interaction which has a $m=1/3$ plateau  is discussed. The phase transition to the classical plateau state is investigated varying the strength of the 4-spin and the third neighbor exchange interaction. The physical interpretation of the mechanism of the quantum plateau  and the relationship to the frustrated period 3 model is given on the basis of the decoupling approximation. The possible universality classes of the transition between quantum and classical plateaus in translationally invariant models are analysed. In the last section, we summarize our results. 

\section{Bosonization  for Translationally Invariant Models}

In order to observe the low energy properties of the above models semi-quantitatively, we employ the bosonization method\cite{ia,sene} which is generally powerful for the description of the one-dimensional spin chains. 
As usual, we first apply the exact Jordan-Wigner transformation to lattice fermions, $\psi_j$, then write the low energy degrees of freedom in terms of left and right-moving continuum fermions with a Fermi wave-vector, $k_Fa=\pi /3$. ($a$ is the lattice spacing). 

\begin{equation}
 \psi_j \approx e^{i\pi j/3}\psi_R(aj)+e^{-i\pi j/3}\psi_L(aj). 
\end{equation}
 We see that under the symmetry transformation of translation by one site:

\begin{equation}
\psi_R\to e^{i\pi /3}\psi_R,\ \ \psi_L\to e^{-i\pi /3}\psi_L. 
\end{equation}

\begin{figure}
\centerline{\includegraphics[width=70mm]{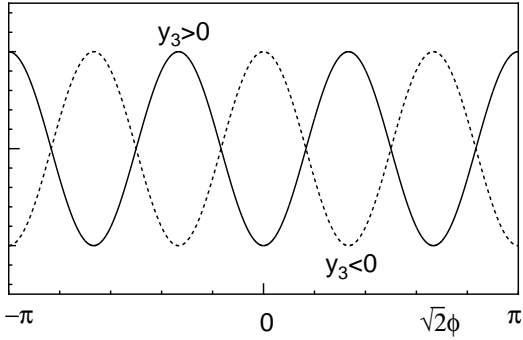}}
\caption{The behavior of the nonlinear term of the bosonized Hamiltonian (\ref{H0bos}).}
\label{pot0}
\end{figure}

This symmetry forbids the usual umklapp term, $\propto (\psi_R^\dagger )^2(\psi_L)^2$, or more accurately and in accord with Fermi statistics: $\propto \psi_R^\dagger \partial_x\psi_R^\dagger \psi_L\partial_x\psi_L$. However the ``triple umklapp''
term $\propto (\psi^\dagger_R)^3(\psi_L)^3$ {\it is} allowed by all symmetries and is expected to occur in the low energy effective Hamiltonian for the uniform frustrated Heisenberg chain (\ref{model1}) and (\ref{fourspin}) at $m=1/3$.  Upon bosonizing, we obtain the sine-Gordon model,
\begin{equation} 
\label{H0bos}
H_0={1\over 2\pi}\int dx \left[K(\pi \Pi)^2+\left({1\over K}\right)
(\partial_x \phi)^2-\frac{y_3}{a^2}\cos 3\sqrt{2}\phi \right],
\end{equation}
where $\phi$ is a boson field, defined on a circle, $\phi \leftrightarrow \phi+\sqrt{2}\pi$, $\Pi$ is the momentum density field conjugate to $\phi$, and  $K$, is the Tomonaga-Luttinger liquid parameter. The spin wave velocity is set equal to unity. A similar expression is also derived in ref. \citen{lo}. The last term in the Hamiltonian, arising from triple umklapp processes,
  is relevant for $K < K_{\c}=4/9$. If this term is relevant, the $Z_3$-symmetry is spontaneously broken and the ground state is 3-fold degenerate as $\phi = 0, \pm {2\pi \over 3\sqrt{2}}$ for $y_3>0$ and $\phi = {\pi\over \sqrt{2}}, \pm {\pi \over 3\sqrt{2}}$ for $y_3 < 0$ as shown in Fig. \ref{pot0}. On the other hand, if the $\cos 3\sqrt{2}\phi$-term is irrelevant, the ground state is the gapless Tomonaga-Luttinger liquid with no degeneracy.

The spin operator $S^z_j$ on the $j$-th site and the dimer operator 
on the $j$-th bond are represented by the boson field $\phi$ as,\cite{ko,ok2}
\bea
\label{sz}
S^z_j-{1 \over 6}&\sim&-C \cos \left(2\pi j/3+ \sqrt{2}\phi \right)\nonumber \\
\v{S}_j\cdot \v{S}_{j+1}
&\sim&\const.+A \cos \left[ \frac{2\pi (j-1)}{3}+\sqrt{2}\phi \right],  
\eea
where  $C$ and $A$  are positive constants
 of the order of unity. Therefore these three ground states correspond to the three degenerate $\uudtype$-type and $\u00type$-type configurations for $y_3 >0$ and $y_3 < 0$ as shown in Fig. \ref{fig1}(a) and (b), respectively. 
Note that for $y_3>0$, the $\phi =0$ ground state has two larger values of $\aver{S^z_j}$ (for $j=3l\pm 1$) 
and one smaller value of $\aver{S^z_j}$ and also two more negative values of $\aver{\v{S}_j\cdot \v{S}_{j+1}}$ 
(for $j=3l$ and $3l-1$) and one less negative value. This is precisely what is expected 
for the classical plateau state of Fig. \ref{fig1}(a): $\aver{\v{S}_j\cdot \v{S}_{j+1}}$ is more negative 
on the bonds joining spins ordered in opposite directions.  Conversely, for $y_3<0$, the $\phi = \pi /\sqrt{2}$ 
ground state has only one larger value of $\aver{S^z_j}$ (for $j=3l$) and one more negative value of
$\aver{\v{S}_j\cdot \v{S}_{j+1}}$ (for $j=3l+1$), corresponding to the quantum plateau state of 
Fig. \ref{fig1}(b). 
Although the sign of $y_3$ cannot be analytically determined, we may conclude $y_3 > 0$ for the Hamiltonian (\ref{model1}) for large enough $\delta$, because  $\udu$ structure is actually observed by Okunishi and Tonegawa\cite{oku}. On the other hand,  the case $y_3 <0$ corresponds to the model (\ref{fourspin}) which we argued exhibits the quantum plateau.

\begin{figure}
\centerline{\includegraphics[width=70mm]{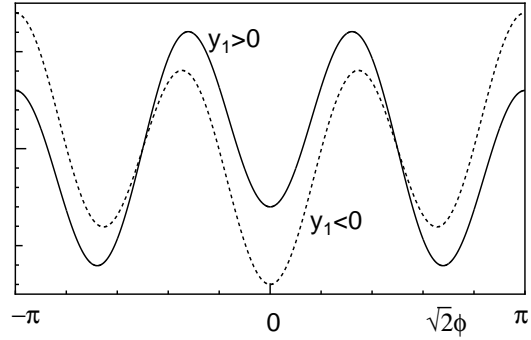}}
\caption{The behavior of the nonlinear terms of the bosonized Hamiltonian (\ref{H0bos})$+$(\ref{H1bos}).}
\label{pot}
\end{figure}

\begin{figure}
\centerline{\includegraphics[width=60mm]{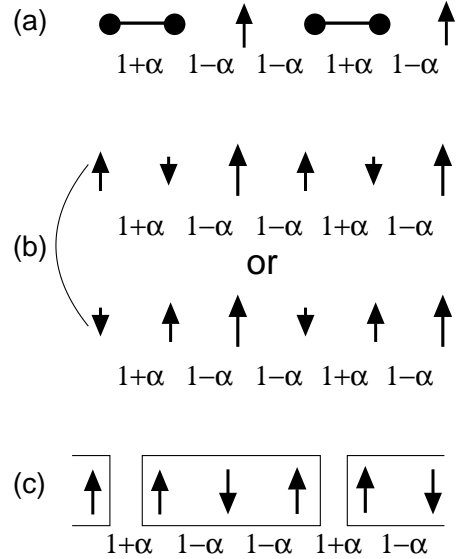}}
\caption{The spin configuration in each phase: (a) $y_1>y_{1c}$, (b) $0<y_1<y_{1c}, (c) y_1<0$.  The pairs of spins connected by thick lines form singlet pairs. The three spins in rectangles form doublets. }
\label{conf0}
\end{figure}
\section{Period-3 Exchange Modulation}
\subsection{Bosonized Hamiltonian}
Now we consider the effect of  period 3 exchange modulation in the models (\ref{model2}) and (\ref{model3}).
Since these models exhibit the classical plateau at $\alpha =0$ and large enough $\delta$, we assume 
that the parameter, $y_3>0$. Exchange modulation is included in the bosonized  
Hamiltonian by the perturbation,\cite{ok,cg,ko}
\begin{equation} \label{H1bos}
H_1={y_1\over 2\pi a^2}\int dx \cos \sqrt{2}\phi,
\end{equation}
where $y_1 = {4\pi\alpha a A}/{3} \propto \alpha$ in terms of the boson field. This term  reduces the 3-fold degeneracy to 2-fold degeneracy for $y_1 > 0$ as long as $|y_1|$ is small as depicted in  Fig. \ref{pot}. For $y_1 < 0$, this term totally removes the degeneracy.

In the former case $y_1 > 0 (\alpha > 0)$, the ground states are  $\phi=\pm\frac{1}{\sqrt{2}}{\rm atan}\left(-\sqrt{\frac{9y_3-y_1}{3y_3+y_1}}\right)$ which are close to $\phi= \pm {2\pi \over 3\sqrt{2}}$ for small $|y_1|$.  Here the branch of arctangent is taken as ${\rm atan}x \in (0,\pi]$. The spin configurations of these two degenerate states for small $|y_1|$ are nearly $\uud$ and $\duu$ according to (\ref{sz}) as shown in  Fig. \ref{conf0}(b). This implies that the $Z_2$ translational symmetry is spontaneously broken in this state. If $y_1$ exceeds $y_{1c}=9y_3$, these two ground states merge to a nondegenerate ground state $\phi=\pi/\sqrt{2}$. This state is the quantum plateau state with $\0u0$ spin configuration shown in  Fig. \ref{conf0}(a) according to the expression (\ref{sz}) which is essentially the same as the ground state of (\ref{model2}) with $\alpha > 0$

In the latter case $y_1 < 0 (\alpha < 0)$, the ground state is $\phi = 0$. According to the expression (\ref{sz}), this phase is the classical plateau state with spin configuration $\udu$ as shown in Fig. \ref{conf0}(c).  In this phase, the three spins coupled via stronger exchange interaction $(1-\alpha)$  form antiferromagnetic trimers whose ground state is the $S=1/2$ doublet $\uudtype$ and $\downarrow\uparrow\downarrow$. In the plateau state, the whole system is covered by the  $\uudtype$ trimers resulting in the classical plateau state with the symmetries of the simple $\udu$ state.

 The transition between the phase with spontaneously broken $Z_2$ symmetry and the $\0u0$-phase is expected to be of the Ising type from symmetry considerations. On the other hand, the transition at $y_1=0$ is a first order transition if $y_3$ is relevant ($\delta > \delta_{\c}$) because the value of $\phi$ jumps at the transition point. If $y_3$ is irrelevant ($\delta < \delta_{\c}$), the transition at $y_1=0$ is a Gaussian transition because the relevant nonlinear term is absent in this case.

\subsection{Renormalization Group Treatment}
For $y_1 <0$, the $\cos \sqrt{2}\phi$-term and $\cos 3\sqrt{2}\phi$-term do not compete. Therefore the ground state is obviously unique $\phi=\frac{\pi}{\sqrt{2}}$ and gapful as long as $\cos \sqrt{2}\phi$-term is relevant. However, for $y_1 > 0$, these two terms compete with each other and drive the Ising type phase transition. To describe this competition, it is necessary to resort to the renormalization group analysis. Up to the second order in $y_1$ and $y_3$, we find the renormalization group equation,
\begin{eqnarray}
\frac{d}{dl}\frac{1}{K}&=& \frac{9}{8}y_3^2, \label{rgk}\\
\frac{dy_3}{dl}&=& (2-\frac{9K}{2})y_3, \label{rg3}\\
\frac{dy_1}{dl}&=& (2-\frac{K}{2})y_1, \label{rg1}
\end{eqnarray}
taking into account the scaling dimensions and the coefficients of the operator product expansion(OPE)  of the operators $(\partial_x \phi)^2$, $\cos \sqrt{2}\phi$ and $\cos 3\sqrt{2}\phi$. Confining ourselves to the region of small $y_1 \propto \alpha$, we have neglected the terms of the order of $y_1^2$. Because we are interested in the behavior near the plateau-non-plateau critical point, we set $K=\frac{4}{9}(1+\frac{1}{2}y_0)$ and keep the lowest order terms in $y_0$. 
\begin{eqnarray}
\frac{dy_0}{dl}&=& -y_3^2, \label{rg0}\\
\frac{dy_3}{dl}&=& -y_0y_3, \label{rg32}\\
\frac{dy_1}{dl}&=& (\frac{16}{9}-\frac{y_0}{9})y_1. \label{rg12}
\end{eqnarray}
Similar set of equations are obtained and analyzed by Kadanoff\cite{kadanoff} and Kitazawa and Nomura\cite{kn} for the double sine-Gordon model with $\cos \sqrt{2}\phi$ and $\cos 2\sqrt{2}\phi$-terms. In the present case, the coupling between $y_3$ and $y_1$ does not appear in the rhs of (\ref{rg12}), because the 
operator $\cos\sqrt{2}\phi$ does not appear in the operator product expansion of 
$\cos\sqrt{2}\phi$ and $\cos 3\sqrt{2}\phi$. 
The first two equations are the well-known Brezinskii-Kosterlitz-Thouless renormalization group equations. The solution which describes the plateau state is given by
\begin{eqnarray}
y_3^2-y_0^2&=&t>0, \\
y_0&=&-\sqrt{t}\tan [\sqrt{t}(l-l_0)],\\
y_3&=&\frac{\sqrt{t}}{\cos [\sqrt{t}(l-l_0)]}.\label{y_3}
\end{eqnarray}
where $t$ measures the distance from the critical separatrix $y_3^2-y_0^2=0$. In the present case $t \propto \delta-\delta_{\c}$. The bare values of $y_{1}$, $y_{0}$ and $y_{3}$  are denoted by $y_{10}\sim \alpha$ and $y_{30}, 
 y_{00} \sim O(1)$.

The integration constant $l_0$ is determined by the initial condition,
\begin{eqnarray}
y_{30}&=&\frac{\sqrt{t}}{\cos [\sqrt{t}l_0]} \sim O(1)
\end{eqnarray}
This implies $\cos \sqrt{t}l_0 \sim O(\sqrt{t})$ and $ l_0 \simeq \pi/2\sqrt{t} + O(1)$ for $t << 1$.
Equation (\ref{rg12}) can be integrated using these solutions to give
\begin{equation}
\ln (y_1/ y_{10})\approx {16l\over 9}+{1\over 9}\ln (y_3/y_{30}).\label{y1}
\end{equation}
for $t << 1$. 
The phase transition takes place by the competition between $y_1$ and $y_3$. Therefore the phase boundary $\alpha_{\c}(\delta)$ can be estimated by setting $y_1 \sim y_3 \sim O(1)$ after enough steps of renormalization, that 
is at some value of $l$.  Since $y_1(l)$ grows exponentially with $l$, from Eq. (\ref{y1}) but 
$y_3(l)$ grows more slowly, from Eq. (\ref{y_3}), becoming O(1) at $l\approx \pi /\sqrt{t}$, we see 
that $y_1(\pi /\sqrt{t})$ must be O(1) on the phase boundary. This implies:
\begin{equation} 
\alpha_{\c} e^{16\pi /9\sqrt{t}}\propto 1
\end{equation}
or
\begin{equation} 
\alpha_{\c} \approx C'e^{-C/\sqrt{\delta -\delta_{\c}}},
\end{equation}
where $C$ and $C'$ are constants of the order of unity.

\subsection{Numerical Diagonalization Studies}

\subsubsection{Ising Transition ($\delta > \delta_{\c}, \alpha=\alpha_{\c}(\delta) >0$)}
To confirm the phase diagram quantitatively, we carry out the numerical diagonalization studies of finite size chains.

\begin{figure}
\centerline{\includegraphics[width=70mm]{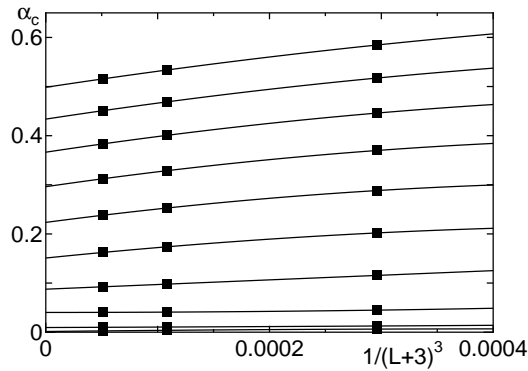}}
\caption{Extrapolation procedure of the finite size critical points $\alpha_{\c}(\delta ; L,L+6)$ to $L \rightarrow \infty$. The dominant size dependence is assumed to be $1/(L+3)^3$. The values of $\delta$ are 1, 0.95, 0.9, 0.85, 0.8, 0.75, 0.7, 0.65, 0.6 and 0.55 from top to bottom.}
\label{prgextra}
\end{figure}

As discussed earlier, the transition between the $Z_2$-symmetry broken intermediate phase and nondegenerate phase for large $\alpha$  belongs to the Ising universality class. The energy gap between the ground state and the lowest excited state with $M=M_s/3$ should behave as $\Delta \propto \const.$ in the non-degenerate phase 
and $\propto \exp(-\const. \times L)$ in the $Z_2$-broken phase.  (The exponentially small 
energy gap for a finite system results from tunnelling between the two degenerate ground states.) On the other hand, at the critical point, $\Delta \propto 1/L$.
 This implies that the phase boundary can be successfully determined by the phenomenological renormalization group (PRG) method.\cite{domb} The finite size critical point $\alpha_{\c}(\delta; L,L+6)$ can be determined by,
\begin{eqnarray}
\lefteqn{L\Delta (\alpha_{\c}(\delta;L,L+6),L)}\nonumber\\
&=&(L+6)\Delta (\alpha_{\c}(\delta;L,L+6),L+6)
\end{eqnarray}
for fixed $\delta$. We take the system size $L$ as multiples of 6 ($L=12, 18, 24$ and 30). The critical point $\alpha_{\c}(\delta)$ in the thermodynamic limit is determined by extrapolating $\alpha_{\c}(\delta; L,L+6)$ to $L \rightarrow \infty$ as shown in Fig. \ref{prgextra} assuming the formula\cite{domb,sakai},
\begin{eqnarray}
\alpha_{\c}(\delta; L,L+6)&\simeq&\alpha_{\c}(\delta)+\frac{\const}{(L+3)^3},
\end{eqnarray}
as plotted in Fig. \ref{phase}. Here we have taken into account that the present transition is expected to belong to Ising universality class from symmetry consideration. 

The exponent $\nu$ of the energy gap $\Delta $ defined by $\Delta  \sim (\alpha_{\c}-\alpha)^{\nu}$ is also estimated by the PRG method as,
\begin{equation}
\nu(L,L+6)=\frac{\ln((L+6)/L)}{\ln\left(\dfrac{(L+6)\Delta '(\alpha_{\c}(\delta; L,L+6),L+6)}{L\Delta '(\alpha_{\c}(\delta; L,L+6),L)}\right)}
\end{equation}
Here $\Delta '$ is the derivative of the energy gap with respect to $\alpha$ at 
fixed $L$. 
Extrapolation to $L \rightarrow \infty$ is carried out assuming the formula\cite{domb,sakai},
\begin{eqnarray}
\nu(L,L+6)&\simeq&\nu+\frac{\const}{(L+3)^2}.
\end{eqnarray}
as plotted in Fig. \ref{prgnu}. Although the size dependence is rather strong, we have verified that the value of $\nu$ is close to the Ising value $(\nu=1)$ as long as the value of $\delta$ is away from $\delta_{\c}$.

In Fig. \ref{phasesc}, $\ln \alpha_{\c}(\delta)$ is plotted against $1/\sqrt{\delta-\delta_{\c}}$ with $\delta_{\c} = 0.487$. This plot shows that the $\delta$-dependence of the numerically obtained value of $\alpha_{\c}$ is well described by the formula $\exp(-\const./\sqrt{|\delta-\delta_{\c}|})$ derived by the renormalization group method in the preceding section.

\subsubsection{First Order Transition ($\delta > \delta_{\c}, \alpha=0$)}

For $\delta > \delta_{\c}$, the numerically obtained ground state energy of the finite size system shows an almost level crossing-like behavior at $\alpha=0$ as shown in Fig. \ref{placr07s} suggesting the first order transition. Although this is not an exact level crossing for finite $L$, the energy eigenvalues per site of two low lying states are almost size-independent near the transition point. Therefore we can interpret these two states as two macroscopically distinct states which are adiabatically connected to the ground states for $\alpha > 0$ and $\alpha <0$ and the first order transition takes place between these two states. This is consistent with the bosonization argument in the preceding section.

\subsubsection{Gaussian Transition ($\delta < \delta_{\c}, \alpha=0$)}
For $\delta < \delta_{\c}$, no level crossing between two macroscopically distinct states takes place suggesting the continuous transition as shown in Fig. \ref{placr03s}.  We have studied the $\pi$-twist 
boundary condition which corresponds to changing the sign of the 
$(S^+_LS^-_1+S^+_1S^-_L)$, $(S^+_{L-1}S^-_1+S^+_1S^-_{L-1})$ and $(S^+_{L}S^-_2+S^+_2S^-_{L})$ terms in the Hamtilonian, finding that the space inversion parity of the ground state reverses at $\alpha =0$.  This is a clear indication of the Gaussian transition driven by the term (\ref{H1bos}) as discussed by Kitazawa\cite{kita} and Kitazawa and Nomura.\cite{kn} We have also numerically checked that the conformal charge $c$ is close to unity  along the line $\alpha=0$ and $\delta <\delta_{\c}$ as shown in Fig. \ref{charge}. Therefore this transition is confirmed to be the Gaussian transition.

\begin{figure}
\centerline{\includegraphics[width=70mm]{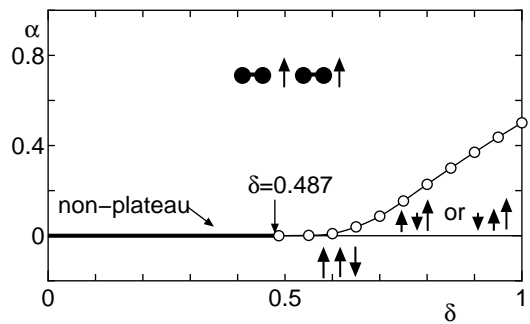}}
\caption{The ground state phase diagram at $m=1/3$.}
\label{phase}
\end{figure}
\begin{figure}
\centerline{\includegraphics[width=70mm]{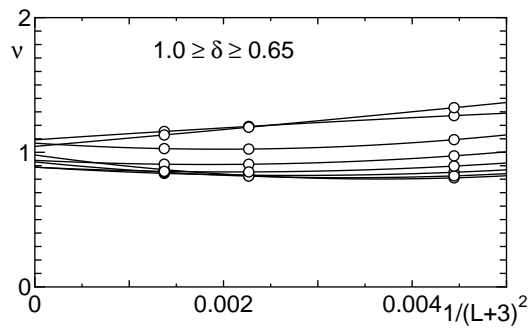}}
\caption{Extrapolation procedure of the finite size critical exponent $\nu(L,L+6)$ to $L \rightarrow \infty$. The dominant size dependence is assumed to be $1/(L+3)^2$.}
\label{prgnu}
\end{figure}
\begin{figure}
\centerline{\includegraphics[width=70mm]{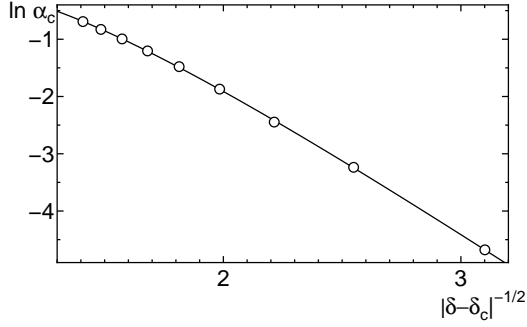}}
\caption{The critical value of $\ln \alpha$ plotted against $1/\sqrt{\delta-\delta_{\c}}$.}
\label{phasesc}
\end{figure}
\begin{figure}
\centerline{\includegraphics[width=70mm]{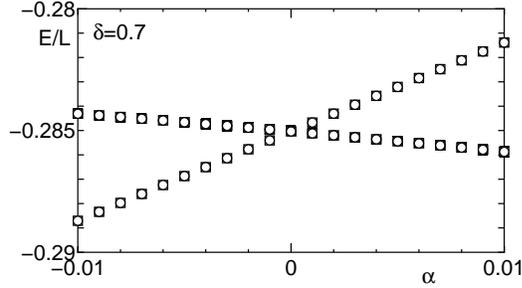}}
\caption{The lowest two energy eigenvalues per site at $\delta=0.7$ for $L=30$(open circles) $L=24$(open squares) and $L=18$ (filled squares). The data points for $L=24$ and $L=18$ are almost covered by those for $L=30$.}
\label{placr07s}
\end{figure}
\begin{figure}
\centerline{\includegraphics[width=70mm]{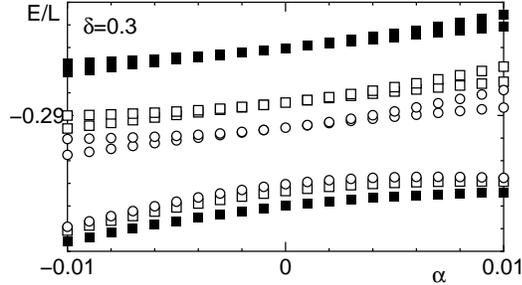}}
\caption{The lowest two energy eigenvalues per site at $\delta=0.3$ for $L=30$(open circles) $L=24$(open squares) and $L=18$ (filled squares). }
\label{placr03s}
\end{figure}
\begin{figure}
\centerline{\includegraphics[width=70mm]{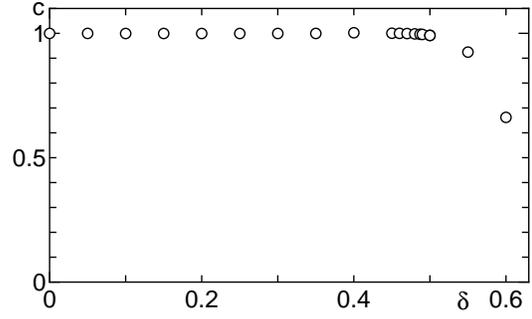}}
\caption{The $\delta$-dependence of the numerically estimated conformal charge for $\alpha=0$. Finite values of $c$ for $\delta > \delta_{\c}$ are numerical artifacts.}
\label{charge}
\end{figure}

\subsubsection{Nature of the $Z_2$-broken phase}

To verify the nature of each phase described in \S 2, we numerically calculate the spin configuration in the plateau state. 

\begin{figure}
\centerline{\includegraphics[width=70mm]{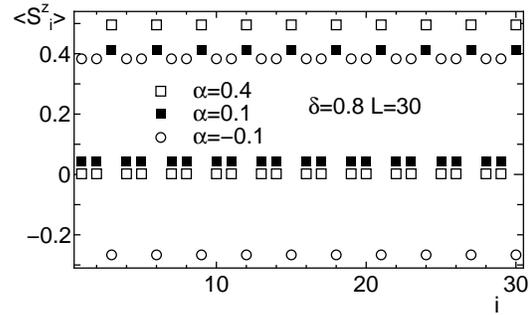}}
\caption{The spatial variation of $\aver{S^z_i}$ in the ground state of the finite chain with $L=30$ for $(\delta,\alpha)=(0.8,-0.1)$(open circles), $(\delta,\alpha)=(0.8,0.1)$(filled squares) and $(0.8,0.1)$(open squares).}
\label{szf0}
\end{figure}

In Fig. \ref{szf0}, we plot the expectation value $\aver{S_i^z}$ for $(\delta,\alpha)=(0.8,0.4)$, $(0.8,0.1)$ and $(0.8,-0.1)$ against $i$  for $L=30$. The spin configuration for $(\delta,\alpha)=(0.8,-0.1)$ shows a clear ${\udu}$ structure. On the other hand, those for $(\delta,\alpha)=(0.8,0.1)$ and $(0.8,0.4)$ look quite similar indicating the $\0u0$ structure at first sight. However, even in the $Z_2$-symmetry broken phase, the ground state of the finite size system does not break the symmetry but is the antisymmetric linear combination of the symmetry broken states $\ket{\uud}$ and $\ket{\duu}$. Therefore it is difficult to distinguish these two phases by  naive numerical calculation for the finite size systems.

\begin{figure}
\centerline{\includegraphics[width=70mm]{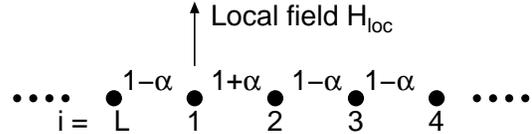}}
\caption{Local field applied to the end spin.}
\label{localf}
\end{figure}

\begin{figure}
\centerline{\includegraphics[width=70mm]{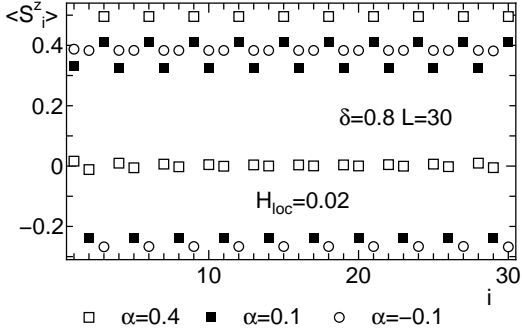}}
\caption{The spatial variation of $\aver{S^z_i}$ in the ground state of the finite chain with $L=30$ for $(\delta,\alpha)=(0.8,-0.1)$(open circles), $(\delta,\alpha)=(0.8,0.1)$(filled squares) and $(0.8,0.4)$(open squares) with small upward local field $H_{\rm loc} =0.02$ at $i=1$ site.}
\label{szf01}
\end{figure}

In order to single out the $Z_2$-symmetry broken phase, we apply the local upward field magnetic field $H_{\rm loc}$ on the $i=1$ site as shown in Fig. \ref{localf}. The calculated expectation values of $\aver{S^z_i}$ are shown in Fig. \ref{szf01}. The $Z_2$-symmetry broken state $\ket{\duu}$ is clearly singled out for $(\delta,\alpha)=(0.8,0.1)$ with weak local field $H_{\rm loc}=0.02$. On the other hand, the ${\0u0}$ spin configuration for $(\delta,\alpha)=(0.8,0.4)$ and the ${\udu}$ spin configuration for $(\delta,\alpha)=(0.8,-0.1)$ are stable against local field except for the tiny polarization. It should be noted that no spatial periodicity is imposed by the local magnetic field in the present calculation.

\begin{figure}
\centerline{\includegraphics[width=70mm]{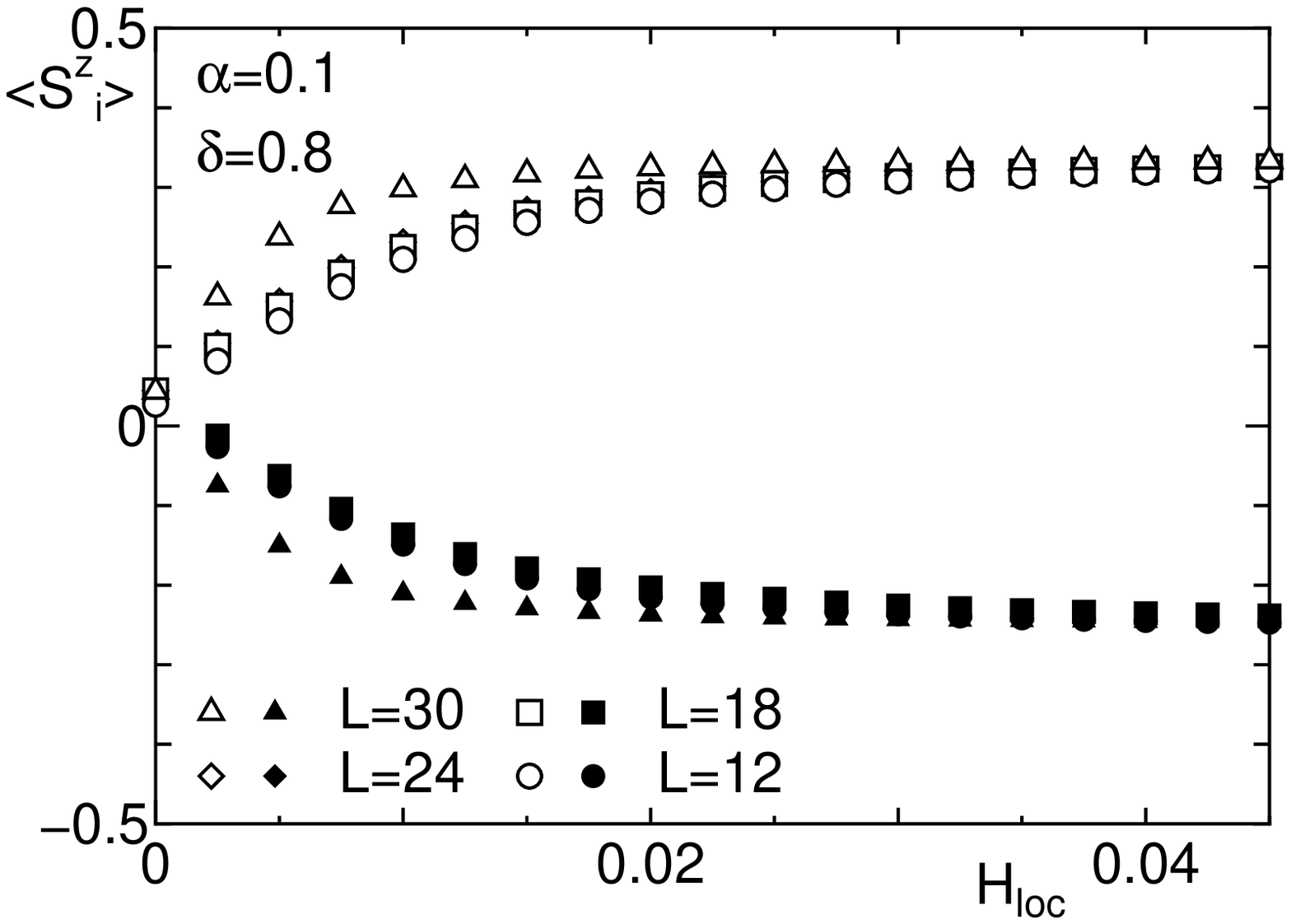}}
\caption{The local field dependence of $\aver{S^z_i}$ on the central sites for $(\delta,\alpha)=(0.8,0.1)$. The filled symbols are $\aver{S^z_{L/2+1}}$ and open symbols are $\aver{S^z_{L/2+2}}$.}
\label{sz0801}
\end{figure}
\begin{figure}
\centerline{\includegraphics[width=70mm]{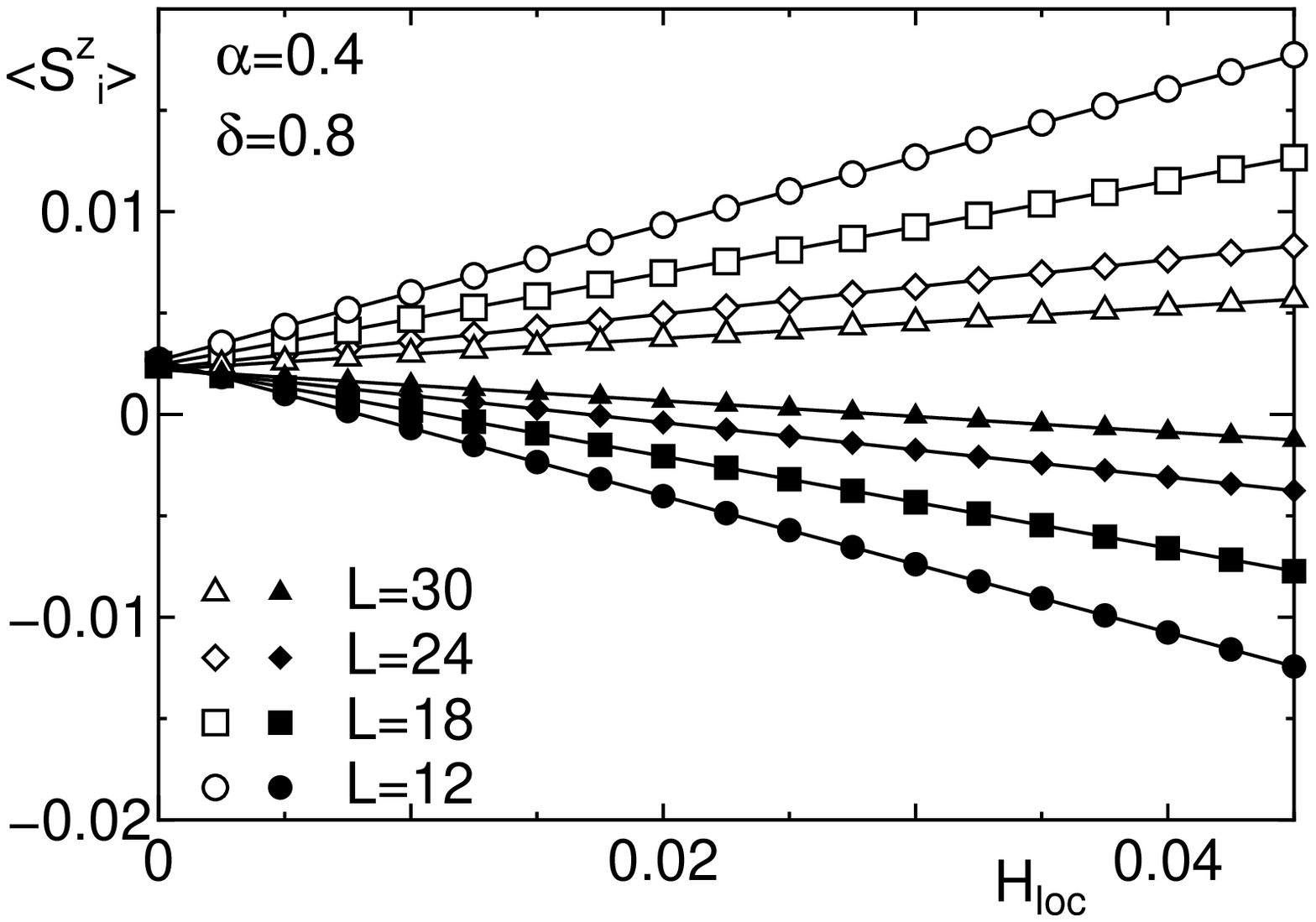}}
\caption{The local field dependence of $\aver{S^z_i}$ on the central sites for $(\delta,\alpha)=(0.8,0.4)$. The filled symbols are $\aver{S^z_{L/2+1}}$ and open symbols are $\aver{S^z_{L/2+2}}$.}
\label{sz0804}
\end{figure}
\begin{figure}
\centerline{\includegraphics[width=70mm]{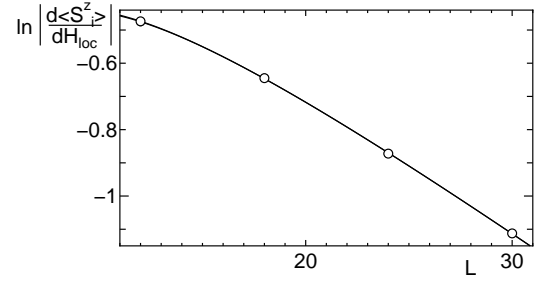}}
\caption{The size dependence of the initial slope $d\aver{S^z_i}/dH_{\rm loc}|_{H_{\rm loc}=0}$ for $(\delta,\alpha)=(0.8,0.4)$. The symbols for $i=L/2+1$ and $i=L/2+2$ overlap within the size of the symbols.}
\label{tanhdep}
\end{figure}

In Figs. \ref{sz0801} and \ref{sz0804}, the local field dependence of the magnetization on the central sites $i=L/2+1$ and $L/2+2$, which are furthest from the position of the local field, are presented for $(\delta,\alpha)=$(0.8,0.1) and (0.8,0.4), respectively.  For $(\delta,\alpha)=(0.8,0.1)$, the magnetization saturates for small local field and this saturation value is insensitive to the system size. This implies these are the bulk values of local magnetization. On the other hand, for $(\delta, \alpha)=(0.8,0.4)$, the magnetization does not saturate for small local field. Fig. \ref{tanhdep} shows the system size dependence of the initial slope of the $H_{\rm loc}$-dependence of $\aver{S^z_{L/2+1}}$ and $\aver{S^z_{L/2+2}}$ for $(\delta,\alpha)=(0.8,0.4)$. It is evident that the slope decreases exponentially with the system size. This confirms that the bulk ground state is actually the $\ket{\0u0}$ state in this phase.

\section{VBS-type Model}

Another way to have a quantum plateau is to reverse the sign of the $\cos3\sqrt{2}\phi$ term in the bosonized language. In this case, the quantum plateau is induced by the spontaneous symmetry breaking in the uniform Hamiltonian. However, within the two-spin exchange interaction we do not find the appropriate model in which quantum plateau is realized. Instead, as explained in section 2, we can construct a translationally invariant VBS-type model (\ref{fourspin}) with  4-spin interaction which exactly shows the 1/3 quantum plateau with spontaneously broken translational invariance.

In terms of the original spin operators $\v{S}_i$, the Hamiltonian (\ref{fourspin}) is rewritten as,
\begin{eqnarray}
\H&=& \frac{1}{24}\sum_{i=1}^L 
[30\v{S}_i\v{S}_{i+1}+20\v{S}_i\v{S}_{i+2}+10\v{S}_i\v{S}_{i+3}]\nonumber\\
&+&8[(\v{S}_i\v{S}_{i+1})(\v{S}_{i+2}\v{S}_{i+3})\nonumber\\
&+&(\v{S}_i\v{S}_{i+2})(\v{S}_{i+1}\v{S}_{i+3})\nonumber\\
&+&(\v{S}_i\v{S}_{i+3})(\v{S}_{i+1}\v{S}_{i+2})]
\label{fourspin2}
\end{eqnarray}
except for the constant term. 

\begin{figure}
\centerline{\includegraphics[width=60mm]{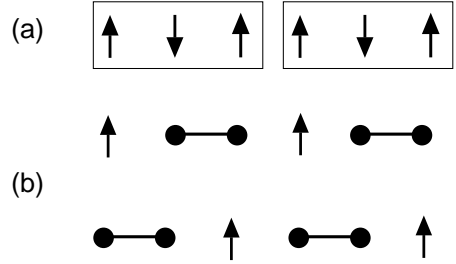}}
\caption{The ground state spin configurations of the Hamiltonian (\ref{four-t}) with open boundary conditions and 6 sites. }
\label{conf4b}
\end{figure}

In order to make clear that this quantum plateau state is a distinct phase from the classical plateau found in model (\ref{model1}), we consider the Hamiltonian which continuously interpolate between the Hamiltonians (\ref{model1}) and (\ref{fourspin2}) as follows,
\begin{eqnarray}
\H&=& \frac{1}{24}\sum_{i=1}^L 
[30\v{S}_i\v{S}_{i+1}+20\v{S}_i\v{S}_{i+2}+10\theta\v{S}_i\v{S}_{i+3}]\nonumber\\
&+&8\theta[(\v{S}_i\v{S}_{i+1})(\v{S}_{i+2}\v{S}_{i+3})\nonumber\\
&+&(\v{S}_i\v{S}_{i+2})(\v{S}_{i+1}\v{S}_{i+3})\nonumber\\
&+&(\v{S}_i\v{S}_{i+3})(\v{S}_{i+1}\v{S}_{i+2})]
\label{four-t}
\end{eqnarray}
and investigate the ground state transition with respect to $\theta$. In the present case, however, the phases on both sides of the transition are phases with spontaneously broken $Z_3$ symmetry. Therefore the lowest energy gap depends on the system size as $\Delta  \sim \exp(-\const. \times L)$. This implies that the naive application of the phenomenological renormalization group analysis is not possible. However, consider the effect of open boundary conditions, for 
a chain of length $L=6n$, in the two phases. 
In the classical plateau phase we expect that open boundary conditions will 
favour a unique plateau state, raising the energy of the other two by an $L$-independent 
amount.  The favoured state is the one drawn in Fig. \ref{conf4b}(a) in 
the case $n=1$, $L=6$.  Note that only $2n-1$ energetically costly 
parallel neighboring spins 
occur in this state.
  The other two classical plateau states have $2n$ parallel spins. On the 
other hand, consider the quantum plateau states with open boundary conditions.  
We expect that the energy will be lowest if no singlet bonds are broken.  In this 
case a singlet bond will occur at one end of the system and a polarized spin 
at the other, as indicated in Fig. \ref{conf4b}(b) for $L=6$. This allows 
two degenerate states, related to each other by space inversion. For a finite open system 
we expect tunnelling process to mix the two low energy quantum plateau states, 
of  Fig. \ref{conf4b}(b) leading to an exponentially low energy excited state. 
On the other hand, in the classical plateau state the gap should be O(1). 

\begin{figure}
\centerline{\includegraphics[width=70mm]{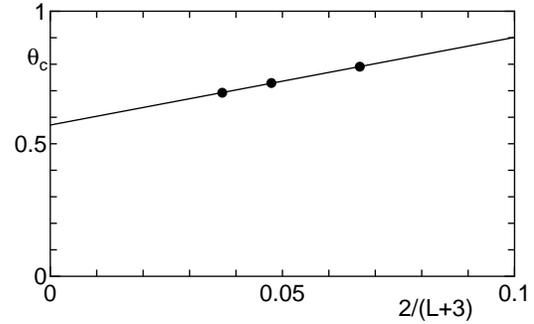}}
\caption{Extrapolation procedure of the finite size critical point $\theta_{\rm c}(L,L+6)$ to $L \rightarrow \infty$.}
\label{qgap20ocr}
\end{figure}

Therefore the energy gap within the $m=1/3$-sector behaves as $\Delta  \sim \const. $ in the former phase and $\Delta  \sim \exp(-\const. L)$ in the latter phase and PRG analysis is possible. The extrapolation procedure of the finite size critical point $\theta_{\rm c}(L,L+6)$ determined from the crossing point of $L\Delta $ for the sizes $L$ and $L+6$ is shown in Fig.\ref{qgap20ocr}. The transition point in the thermodynamic limit is $\theta_{\rm c}\simeq 0.57$. Although the precise value of the critical point is not accurate enough due to the relatively large size dependence of finite size critical point, this analysis clearly shows that the $\u00type$ and $\uudtype$ states are two distinct phases even in the uniform chain.

This transition can be interpreted in terms of the bosonized language. At $\theta=\theta_{\rm c}$, the coefficient $y_3$ vanishes. Therefore, the higher order umklapp term
\begin{equation}
H=\frac{y_6}{2\pi a^2} \int dx \cos 6\sqrt{2}\phi
\end{equation}
comes into play. This term is irrelevant or relevant depending on whether $ K >1/9$ or $K < 1/9$. If this term is irrelevant, the transition is expected to be the second order Gaussian transition with conformal charge $c=1$. If this term is relevant and $y_6 <0$, the ground state is 6-fold degenerate at the transition point. This corresponds to coexisting classical and quantum plateaus.  Turning on $y_3$, with either sign, either the classical or quantum plateau states are favored. In this case, the transition is a first order transition and level crossing behavior is expected. If this term is relevant and $y_6 > 0$, the $y_3$ and $y_6$ terms compete near the point $y_3=0$ and two Ising type transitions should take place as discussed in the competition between $\cos \sqrt{2}\phi$ and  $\cos 2\sqrt{2}\phi$ terms in the Ashkin-Teller model.\cite{kadanoff,kn} 

Numerically we find only a single transition and the smallest energy gap $\Delta$ with periodic boundary conditions scales as $L\Delta \sim {\rm const.}$ near the critical point. We also find the conformal charge $c\simeq 1.0$  extrapolating from $L=12,18$ and 24. Therefore we conclude that the first one of the above three senarios is valid in the present case and the transition is a Gaussian transition. In order to confirm that $y_6$ is irrelevant and $y_3$ is relevant near the transition point, we determine $K$ using the relationship between the excitation spectrum and $K$. Here we employ the relation which contains the lowest excitation energy of the mode with wave number $k=2\pi/3$,
\begin{equation}
\frac{K}{2}=\frac{L\Delta (k=2\pi/3, M=M_s/3)}{2\pi v_{\rm s}}
\end{equation}
where $v_{\rm s}$ is the spin wave velocity given by,
\begin{equation}
v_{\rm s}=\frac{L\Delta (k=2\pi/L, M=M_s/3)}{2\pi}.
\end{equation}
Extrapolating the value of $K(L)$ estimated for finite size $L$ to $L \rightarrow \infty$ as $K(L) \simeq K+C/L^2$  from $L=12,18$ and 24, we find $K \simeq 0.368$ which is between $4/9$ and $1/9$. Therefore $y_3$ is relevant and $y_6$ is irrelevant and the condition for the Gaussian transition is fulfilled. 

\begin{figure}
\centerline{\includegraphics[width=70mm]{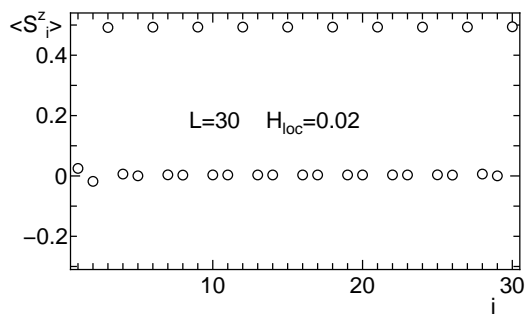}}
\caption{The spatial variation of $\aver{S^z_i}$ in the ground state of the decoupled model $H^{\rm MF}$ with $L=30$ with small upward local field $H_{\rm loc} =0.02$ at $i=1$ site.}
\label{szmf}
\end{figure}

To understand the physical origin of the $\u00type$ phase in the 4-spin Hamiltonian, we decouple the 4-spin terms in (\ref{fourspin2}) as,
\begin{eqnarray}
\lefteqn{(\v{S}_i\v{S}_{j})(\v{S}_{k}\v{S}_{l}) \rightarrow \aver{(\v{S}_i\v{S}_{j})}(\v{S}_{k}\v{S}_{l})}\nonumber\\
&+& (\v{S}_i\v{S}_{j})\aver{(\v{S}_{k}\v{S}_{l})}-  \aver{(\v{S}_i\v{S}_{j})}\aver{(\v{S}_{k}\v{S}_{l})}
\end{eqnarray}
Assuming the complete $\0u0$ state, we take 
\begin{eqnarray}
\aver{(\v{S}_{3l+1}\v{S}_{3l+2})}&=&-\frac{3}{4}\nonumber\\
\aver{(\v{S}_{3l}\v{S}_{3l+3})}&=&\frac{1}{4}\nonumber\\
\aver{(\v{S}_{i}\v{S}_{j})}&=&0 \ \ \mbox{other terms}
\label{mfval}
\end{eqnarray}
Then we have
\begin{eqnarray}
\H^{\rm MF}&=& \frac{1}{24}\sum_{l=1}^{L/3} 
[24(\v{S}_{3l-1}\v{S}_{3l}+\v{S}_{3l}\v{S}_{3l+1})\nonumber\\
&+&32\v{S}_{3l+1}\v{S}_{3l+2}\nonumber\\
&+&20(\v{S}_{3l-1}\v{S}_{3l+1}+\v{S}_{3l}\v{S}_{3l+2}+\v{S}_{3l-2}\v{S}_{3l})\nonumber\\
&+&10(\v{S}_{3l-2}\v{S}_{3l+1}+\v{S}_{3l-1}\v{S}_{3l+2})\nonumber\\
&+&4\v{S}_{3l}\v{S}_{3l+3}]
\label{mfham}
\end{eqnarray}
except for the constant terms. If the last term is neglected, this Hamiltonian is equivalent to the Hamiltonian (\ref{model3}) with $\alpha=\frac{1}{7}\simeq 0.143$ and  $\delta=\frac{5}{7}\simeq 0.714$ which belongs to the $\u00type$ phase in the phase diagram Fig. \ref{phase}. 
To demonstrate that the last term is not harmful to the $\0u0$ structure of the plateau, we carried out the numerical diagonalization of the Hamiltonian (\ref{mfham}) and calculated the local spin configuration in the ground state with local field at $i=1$. The result is shown in Fig. \ref{szmf} for $L=30$. The expectation values of the dimer spins almost vanishes and the polarized spins are almost equal to $1/2$. 
\begin{figure}
\centerline{\includegraphics[width=70mm]{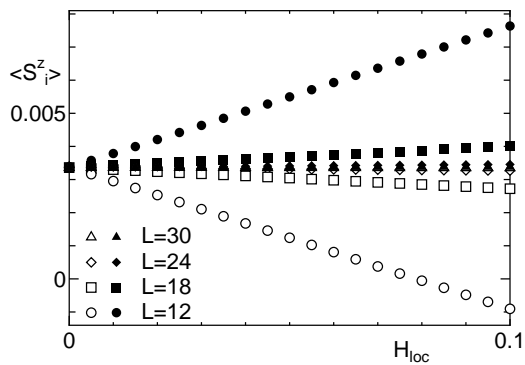}}
\caption{The local field dependence of $\aver{S^z_i}$ on the central sites in the ground state of the decoupled model $H^{\rm MF}$. The filled symbols are $\aver{S^z_{L/2+1}}$ and open symbols are $\aver{S^z_{L/2+2}}$.}
\label{sz4nloc}
\end{figure}
We also calculated the spin-spin correlation functions which we have assumed in Eq. (\ref{mfval}) as,
\begin{eqnarray}
\aver{(\v{S}_{3l+1}\v{S}_{3l+2})}&=&-0.74294967\nonumber\\
\aver{(\v{S}_{3l}\v{S}_{3l+3})}&=&0.24014714  \nonumber\\
|\aver{(\v{S}_{i}\v{S}_{j})}|&\leq&0.04020856 \ \ \mbox{other terms}
\end{eqnarray}
for $L=30$. The results agree with the assumed set of values (\ref{mfval}) quite well. Therefore we may conclude our assumptions (\ref{mfval}) are self-consistent and we can describe the physical picture of the quantum plateau state in the following way. 

 The periodic array of dimers induce the effective period 3 exchange modulation via the 4-spin term self-consistently, resulting in the quantum magnetization plateau with spontaneous $Z_3$-translational symmetry breakdown at $m=1/3$. This implies the 4-spin interaction can be generally the source of dimer type order in low dimensional magnets.

\section{Summary}

The competition between the $m=1/3$ classical and quantum magnetization plateau states in the $S=1/2$ frustrated Heisenberg chains  with space inversion invariant period 3 exchange modulation and 4-body interaction  is investigated by the renormalization group and numerical diagonalization method.  

 The conventional $S=1/2$ frustrated Heisenberg chains is known to exhibit the 3-fold degenerate $\uudtype$-type classical plateau accompanied by the spontaneous $Z_3$ translational symmetry breakdown. Two of them ($\ket{\uud}$ and $\ket{\duu}$) turned out to be robust against the period 3 exchange modulation which favors the $\0u0$ phase up to a critical value of the modulation amplitude ($\sing = $  singlet dimer) resulting in the $Z_2$ translational symmetry broken phase. Another $\uudtype$-type state $\ket{\udu}$  is stabilized for period 3 modulation with opposite sign. The transition between the $\0u0$-phase and $Z_2$-broken phase is the Ising transition and that between the $\udu$-phase and $Z_2$ broken phase is the first order transition. The spin configuration in each phase is numerically verified. 

A translationally invariant Hamiltonian with the exact $\u00type$-type quantum plateau ground state is also presented based on the VBS-type construction. The spontaneous $Z_3$ translational symmetry breakdown takes place in the ground state. The phase transition from the $\u00type$-type phase to the  $\uudtype$-type phase takes place as the strength of the 4-spin term and third neighbor interaction is varied. This implies that these two phases are distinct phases. The universality class of this transition is discussed in bosonized language. From numerical analysis, this transition turned out to be the Guassian transition. A physical picture of the quantum plateau phase is presented based on the decoupling approximation.

The authors thank T. Tonegawa and K. Okunishi for stimulating discussion. The research of KH  is supported by a Grant-in-Aid for Scientific Research from the Ministry of Education, Culture, Sports, Science and Technology, Japan. IA thanks the Tokyo Institute of Technology for hospitality.  The research of IA was partly supported by JSPS, NSERC of Canada and the Canadian Institute for Advanced Research. The numerical diagonalization is carried out using the package TITPACK ver.2 coded by H. Nishimori. The numerical computation in this work has been carried out using the facilities of the Supercomputer Center, Institute for Solid State Physics, University of Tokyo and the Information Processing Center, Saitama University.

\end{document}